# Additively manufactured hybrid anisotropic pentamode metamaterials

**Kaivan Mohammadi[1], Mohammad R. Movahhedy[1], Igor Shishkovsky[2], Reza Hedayati[3,*]**

[1]Additive Manufacturing Lab (AML), Department of Mechanical Engineering, Sharif University of Technology, Azadi Avenue, 11365-11155 Tehran, Iran

[2]Center for Design, Manufacturing and Materials, Skolkovo Institute of Science and Technology, 3 Nobel Str., 121205 Moscow, Russia

[3]Novel Aerospace Materials Group, Faculty of Aerospace Engineering, Delft University of Technology (TU Delft), Kluyverweg 1, 2629 HS Delft, the Netherlands

---

[*] Corresponding author, email addresses: r.hedayati@tudelft.nl, rezahedayati@gmail.com,

# Abstract

Pentamode metamaterials are a type of extremal designer metamaterials which are able to demonstrate high rigidity in one direction and extremely high compliance in other directions. Pentamodes can, therefore, be considered as building blocks of exotic materials with any arbitrarily selected thermodynamically admissible elasticity tensor. The pentamode lattices can then be envisioned to be combined to construct intermediate extremal materials such as quadramodes, trimodes, and bimodes. In this study, we constructed several primary types of anisotropic pentamode lattices (with midpoint positioning of 10%, 15%, 20%, 25%, 30%, 35%, and 42% of the main unit cell diagonal) and then combined them mutually to explore the dependence of elastic properties of hybrid pentamodes on those of individual constructing lattices. Several anisotropic individual and hybrid pentamode lattice structures were produced using MultiJet Additive Manufacturing technique and then mechanically tested under compression. Finite element models were also created using COMSOL Multiphysics package. Two-component hybrid pentamode lattices composed of individual lattices with extensively different (as large as of 2 order of magnitudes) $B/G$ ratios were constructed and analysed. It was demonstrated that it is possible to design and construct composite intermediate extremal materials with arbitrary eigen values in the elastic tensor. It is concluded that the elastic $E$, shear $G$, and bulk moduli $B$ of the hybrid structure are the superpositions of the corresponding moduli of the individual lattice structures. The Poisson's ratio $\nu$ of the hybrid pentamode structure equals that of individual structure with higher Poisson's ratio. The yield stress $\sigma_y$ of the hybrid pentamode lattice structure depends on the elastic moduli of the constructing lattice structures as well as yield stress of the weaker lattice structure.

Mechanical metamaterials are rationally designed lattices that demonstrate exotic mechanical properties such as negative compressibility [1], negative Poisson's ratio [2], double negativity [3], and fluid-like behaviour [4-7] which are not found or are very rare in nature. The latter, commonly known as pentamode metamaterials, are a type of extremal designer metamaterials [5, 8] which are able to demonstrate high rigidity in one direction and extremely high compliance in other directions. This has led to proposition of pentamodes having shear moduli orders of magnitude larger than their bulk moduli [9]. This has earned pentamodes the name "metafluids" as fluids also have very small shear resistivity while they demonstrate very high incompressibility levels [7]. However, unlike fluids, pentamodes have the capability of being programmed into having characteristics such as stability, inhomogeneity, and anisotropy [10].

Pentamode metamaterials were first introduced theoretically by Milton and Cherkaev in 1995 [11], but they were not realized experimentally until very recently [9] thanks to extensive developments in Additive manufacturing (AM) technologies. Mathematically speaking, pentamodes are materials that have one extremely high pressure-type eigenvalue in their elasticity tensor, while the values of other eigenvalues in the elasticity tensor are negligible. Pentamodes can, therefore, be considered as building blocks of exotic materials with any arbitrarily selected thermodynamically admissible elasticity tensor. The pentamode lattices can then be envisioned to be combined to construct intermediate extremal materials such as quadramodes, trimodes, and bimodes [11]. Even though several designs have been proposed for pentamodes [12-15], the most common pentamode metamaterials are face-centre cubic (fcc) diamond-shaped lattice structures made up of double-cone linkages, the cones attached to each other at their larger diameter $D$ at the midpoint of the linkage [6, 16-20]. If the smaller diameter of the Pentamode is chosen small enough (i.e. $d/a < 4\%$ [5, 7]), the effect of the larger diameter size on the elastic modulus becomes negligible and the smaller diameter determines the macro-scale properties of the structure [7].

In their early work, Milton and Cherkaev [11] suggested to create holes in the linkages or to convert vertices into a cluster of vertices to avoid interpenetration of individual Pentamode lattices which are used to create intermediate extremal materials. Another approach to avoid intersection of linkages is to combine pentamodes with anisotropic microgeometries. In that way, while the microgeometry of each individual pentamode lattices is chosen to accommodate to its corresponding pre-defined elasticity tensor, both lattices can be combined in such a way that they do not intersect with each other, therefore maintaining their initial design intact. One way of making a pentamode metamaterial anisotropic is to move the midpoint P at initial position of $P = (0.75\,a, 0.75\,a, 0.25\,a)$ along the main diagonal of the unit cell (Figure 1d) in such a way that its initial distance from the vertex is changed from $0.25\,L$ to any other distance in the range of $0$ and $0.5\,L$, where $L = \sqrt{3}a$ is the length of the main diagonal of the unit cell.

While anisotropy in design of pentamodes have been previously studies acoustically in [4, 21], making composites out of individual pentamode lattices and the effect of elastic properties such as elastic modulus, yield stress, and Poisson's ratio of the primary individual lattices on the elastic properties of the hybrid structure has never been studies before. In this study, we aim to construct several types anisotropic pentamodes and then to combine them mutually to explore the dependence of elastic properties of hybrid pentamodes on individual constructing pentamode lattices.

In this study, the small diameter $d = 300\,\mu m$ and the large diameter $D = 1100\,\mu m$ was used in all the models. The unit cell size was also kept constant $a = 8$ mm. We chose seven particular positions for P with distances from the vertex of $0.1\,L$, $0.15\,L$, $0.2\,L$, $0.25\,L$, $0.3\,L$, $0.35\,L$, and $0.42\,L$ (Figure 1). Finite element (FE) models were created using COMSOL Multiphysics package (Stockholm, Sweden). All the FE models were lattices made up of 5×5×5=125 unit cells, as demonstrated in Figure 1. The struts were discretised using around $10^6$ tetrahedral elements. A mesh sensitivity analysis was performed and the numerical result converged for element sizes smaller than $20\,\mu m$ at diameter $d$ and at $140\,\mu m$ at diameter $D$ (see Figure A1-A2 in the Appendix for further information). MUMPS static solver in COMSOL was used to solve continuum mechanics equations. Linear elastic material model with mechanical properties of the constituent polymeric material ($E_s = 1.46$ GPa, $\sigma_{ys} = 43$ MPa, $\nu_s = 0.4$, $\rho_s = 1020\,kg/m^3$) was implemented, and a stepwise (in 10 steps) uniform displacement of 1010 $\mu m$ was applied on the top side of the lattice structure. The lowermost nodes of the lattice structure were constrained in all the directions and the uppermost nodes of the lattice structure were only allowed to move in the loading direction.

Three of the anisotropic pentamode designs with midpoint positionings of $0.15\,L$, $0.25\,L$, and $0.42\,L$ were selected for AM and mechanical testing. All the specimens were fabricated using Projet 3500 HD Max 3D printer (3D SYSTEMS, SC, US). 3D SYSTEMS' proprietary VisiJet M3 Crystal polymer was used for manufacturing the body of the lattice structure and pure VisiJet S300 wax was used as the support material to hold overhanging parts, which was removed after printing process by heating the samples up to $60\,°C$. In addition, a hybrid pentamode metamaterial with two constituting pentamode lattices with midpoint positionings of $0.15\,L$ and $0.42\,L$ was also manufactured using the same manufacturing parameters. In order to avoid free movement of the two lattices with respect to each other, the two lattices were softly connected to each other by means of two very thin ($200\ \mu m$ in diameter) bars printed on two opposite edges of the lower-most surface of the specimen.

An INSTRON 5969 servo mechanical testing machine with an INSTRON 2580-500N load cell was used to load the specimens under uniaxial compressive loads with a displacement rate of 5 mm/min. The elastic properties of the structures (elastic modulus, yield stress, and Poisson's ratio) were calculated from the obtained linear regime of load-displacement curves in accordance to the ISO standard 13314:2011. The properties were then normalized using the mechanical properties of the constituent material. The stress-strain curves of all the lattices are provided in Figure A3 of the Appendix. Relative density is defined as the ratio of the density of a lattice structure to that of the constituent material it is made of. Experimental relative density was measured by dry-weighing method, and the numerical relative density was calculated by dividing the volume occupied by constituent material in a unit cell to the volume of the whole unit cell.

All the physical and mechanical properties obtained from experimental tests and numerical models are in good agreement with each other except for the case of 42% L. Other than that point, the maximum differences between numerical and experimental values for relative density, elastic modulus, Poisson's ratio, and yield stress are 10.6%, 8%, 6.6%, and 12.6%, respectively (Figure 4). The numerical/experimental discrepancies for the case of 42% L for the noted physical/mechanical properties are respectively 16.7%, 2.6%, 29.9%, and 50.7%, (Figure 4).

As expected, the variation of relative density $\mu$ with respect to its diagonal location is almost symmetrical and varies in a small range ($0.046 < \mu < 0.054$). As seen in Figure 4b,e, by moving the midpoint from the vertex (10% L) towards the centroid (42% L) of the unit cell, there is a continuous decrease in normalized elastic modulus (from $3.25 \times 10^{-4}$ to $1.26 \times 10^{-4}$, i.e. 61% decrease), normalized shear modulus (from $3.79 \times 10^{-4}$ to $1.38 \times 10^{-4}$, i.e. 64% decrease), and Poisson's ratio (from 0.59 to 0.37, i.e. 37% decrease). Stress analysis in the struts demonstrated that as the midpoint position changes from the vertex towards the centroid, the main type of load sustained by the struts changes from compression to bending which contributes to decrease in the elastic modulus. For the case of yield stress, the normalized yield stress first increases from $0.95 \times 10^{-4}$ (at 10% L) to a peak of $1.22 \times 10^{-4}$ (at 25% L), and then it decreases to the minimum value of $0.68 \times 10^{-4}$ when very close to the centroid (at 42% L). Similarly, normalized bulk modulus has a bell-shaped curved with a peak of $2.1 \times 10^{-3}$ at 25% L and minimum value of $1.1 \times 10^{-3}$ at 42% L (Figure 4b).

Accordingly, the ratio $B/G$ (also known as figures-of-merit, FOM) has a bell-shaped curve with peak point of 50 at 25% L and minimum of ≈3 at 10% L and 42% L (Figure 4c).

As for the case of hybrid pentamode metamaterial, the differences between the numerical and experimental values of relative density, elastic modulus, Poisson's ratio, and yield stress for the hybrid pentamode were 22.1%, 2.7%, 8.1%, and 38.8%, respectively (experimental results being: $\mu = 0.131$, $E = 545$ kPa, $\nu = 0.507$, and $\sigma_y = 12.87$ kPa). The relative density, elastic modulus, Poisson's ratio, and yield stress and values of FE hybrid models constructed from individual lattice structures with different %L is demonstrated in Figure 5 and the values are listed in Table A1 in the Appendix. As for the relative density, it was expected that the relative density of the hybrid structure is the summation of the relative density of each individual lattice structures. This is confirmed since the expected and numerical curves have very similar topologies and both are in the range of 0.0921 and 0.1079 (Figure 5a and Figure A6a in Appendix). As for the elastic, shear, and bulk moduli, a similar superposition hypothesis was expected as the stiffness of the whole hybrid structure originates from the stiffness of the individual structures and as a result, the hybrid structure can be considered as two separate structures working in parallel, and thus each taking part of the load. The hypothesis was confirmed by numerical results and as it can be observed in Figure 5b-d and Figure A6b-d in Appendix, the expected surfaces correlate well with the corresponding numerical results for elastic, shear, and bulk moduli.

As for the yield stress, it is expected that the yield stress of the hybrid structure is determined by the load which causes the weaker structure reach its maximum load. If two structures I (compliant) and II (stiff) with elastic modulus and yield stress of $E_I$, $\sigma_{y,I}$, and $E_{II}$, $\sigma_{y,II}$ are combined, the ratio of the applied load sustained by the more compliant structure is $\frac{E_I}{(E_I+E_{II})}F$, where $F$ is total external load applied on the structure. Therefore, the maximum stress in the compliant component in the absence and presence of the second stiff component II is $\sigma_{max,I}$ and $\frac{E_I}{(E_I+E_{II})}\sigma_{max,I}$, respectively. Therefore, the yield structure of the hybrid structure can be considered as $\sigma_y = \frac{(E_I+E_{II})}{E_I}\sigma_{y,I}$. The yield strength curve obtained from this hypothesis and the numerical yield strength curve are demonstrated in Figure 5d and Figure A6d in Appendix, and as it can be seen they have very similar topologies. The expected results are in the range of 5.58-10.20 kPa and the numerical results are in the range of 5.28-9.8 kPa which are very close to each other.

As for the Poisson's ratio, as long as the two individual structures do not interact with each other, the lattice with higher Poisson's ratio expands larger. As the imposed strain on both individual structures is identical (and equal to that of the hybrid structure), it is expected (and later confirmed numerically) that the Poisson's ratio of the hybrid structure is determined by the Poisson's ratio of the individual structure with higher Poisson's ratio value. Comparison of the expected and numerical results shows their good accordance in both topology shape and range (i.e. $0.36 < \nu < 0.63$).

In summary, the effect of varying the midpoint positioning of the pentamode structure (as a way of making pentamode anisotropic) on its mechanical properties was investigated numerically and experimentally. Moving the midpoint point along the main diagonal from the vertex to the centroid of the unit cell decreases elastic modulus, shear modulus, and Poisson's ratio almost steadily. The yield stress and bulk modulus have their peaks at the midpoint original position (i.e. %25 L). Hybrid intermediate pentamode lattice structures were also constructed from individual pentamode lattice structures with different B/G ratios (with difference in B/G as large as 2 orders of magnitude). First and foremost, it was demonstrated that it is possible to design and construct composite intermediate extremal pentamode materials with arbitrary eigen values in the elastic tensor. Furthermore, it was concluded that the relative density, elastic modulus, shear modulus, and bulk modulus of the hybrid structure can be determined by superposition of the noted properties of the individual lattice structures. The results also showed that Poisson's ratio of the hybrid pentamode structure equals that of individual structure with higher Poisson's ratio. Moreover, the yield stress of the hybrid pentamode lattice structure depends on the elastic moduli of the constructing lattice structures as well as yield stress of the weaker lattice structure.

The data that support the findings of this study are available from the corresponding author upon reasonable request.

**Figures**

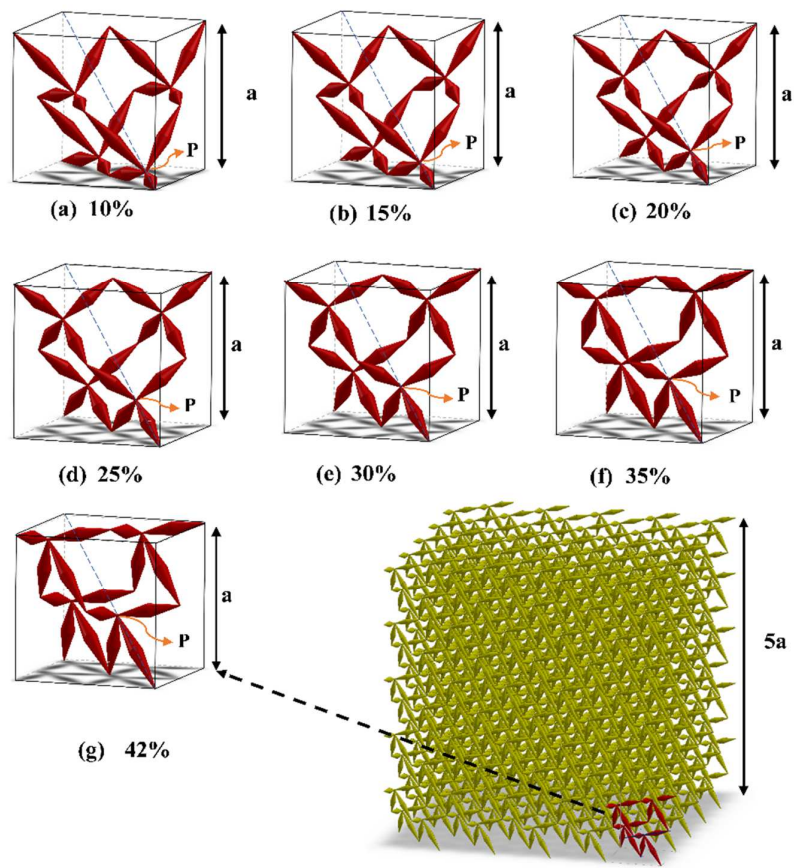

*Figure 1: Anisotropic unit cells of pentamode metamaterial with P positioning from 10-42%*

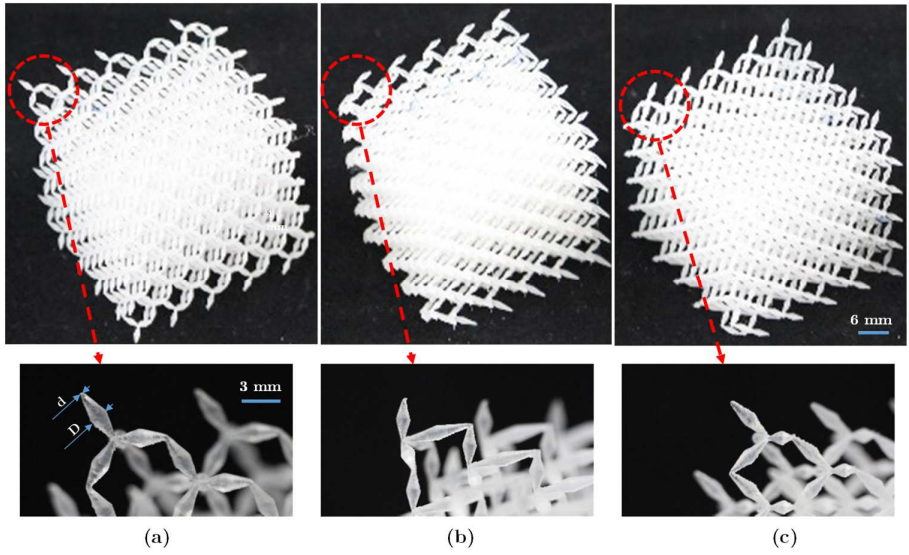

*Figure 2: Images of the specimens manufactured with positioning of midpoint (a) P= 25%, (b) P= 42%, and (c) P=15%*

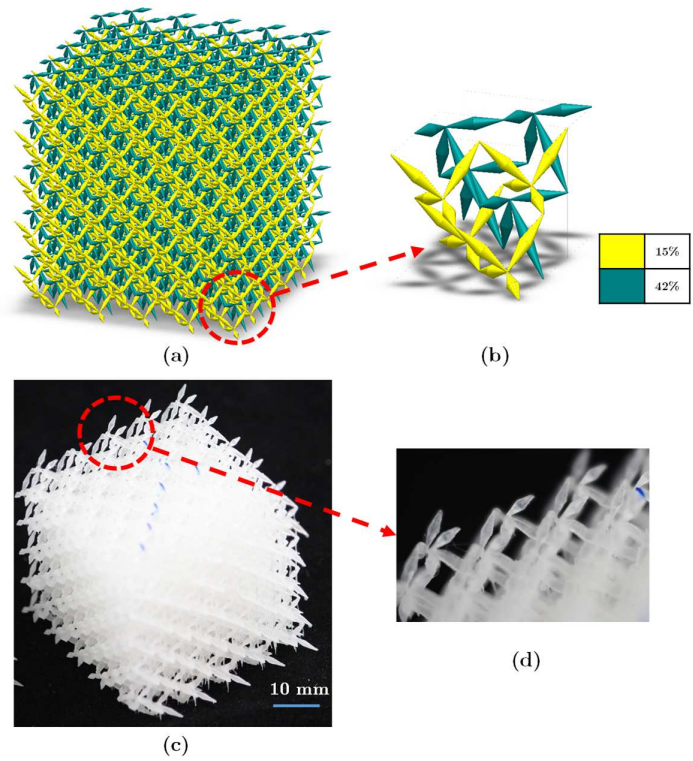

*Figure 3: (a) Lattice and (b) a unit cell of a structure combined of two pentamode metamaterials with 15% and 42% midpoint positioning. (c-d) The manufactured hybrid pentamode metamaterial*

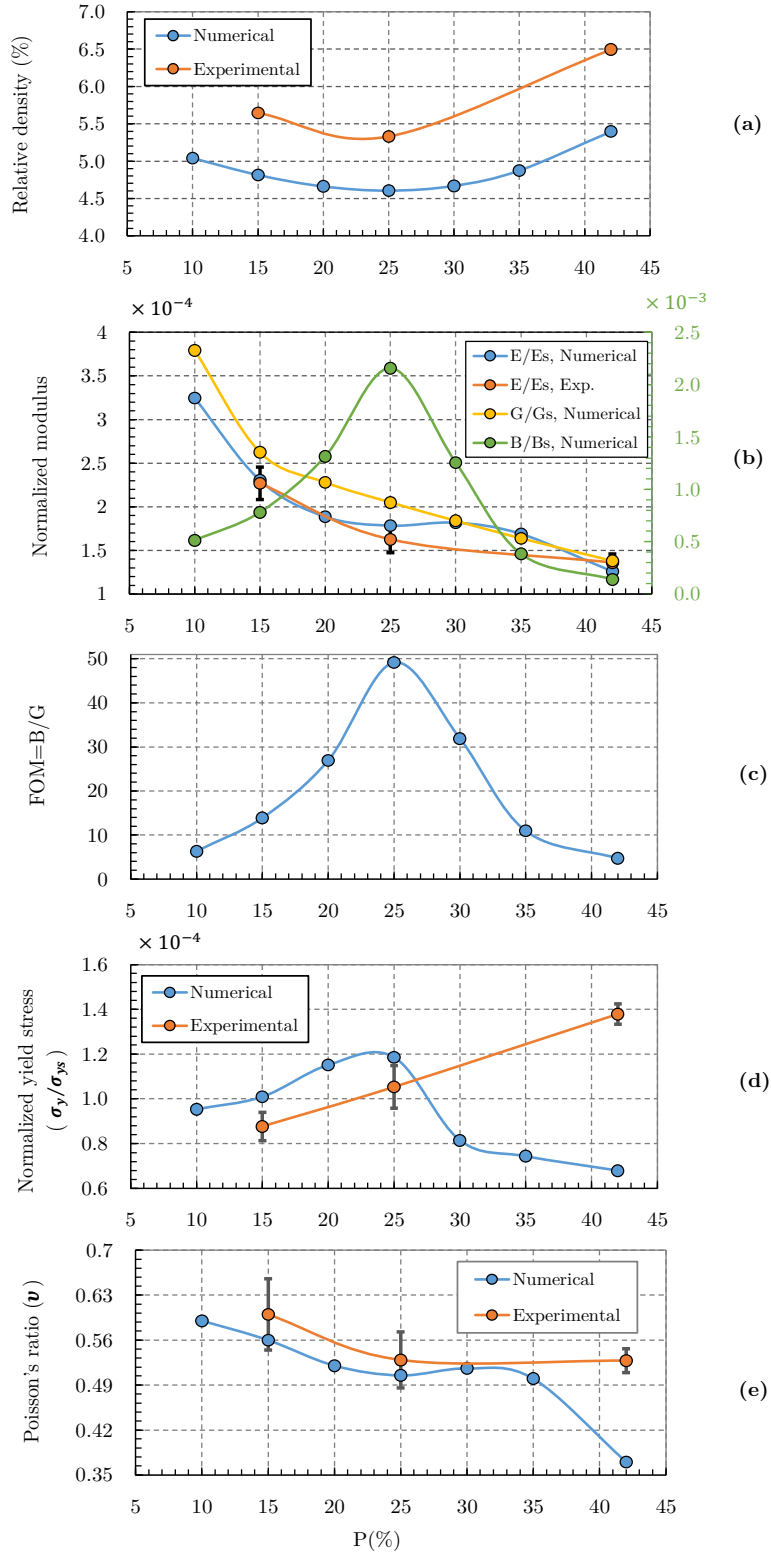

Figure 4: Variation of (a) relative density, (b) normalized modulus, (c) FOM, (d) normalized yield stress, and (d) Poisson's ratio with respect to midpoint positioning.

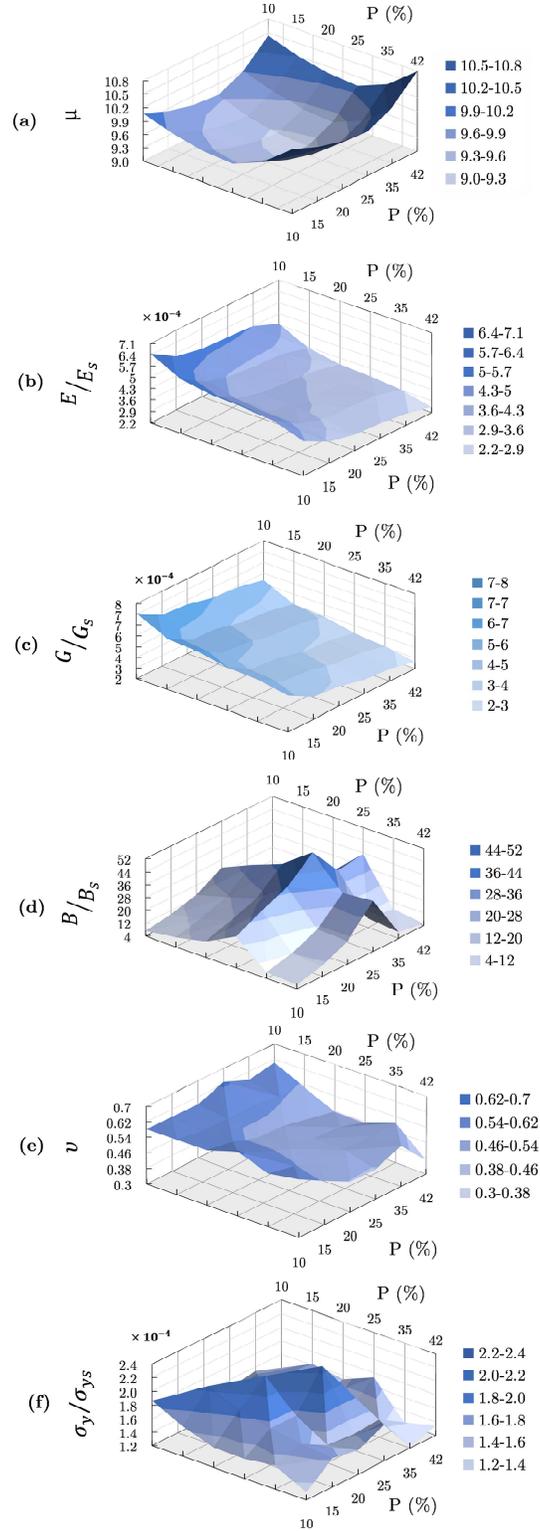

*Figure 5: Variation of (a) relative density, (b) elastic modulus, (c) shear modulus, (d) bulk modulus (e) Poisson's ratio, and (f) yield stress of hybrid Pentamode metamaterials depending on the corresponding properties of constructing individual lattices.*